\documentclass[twocolumn,prb,showpacs,preprintnumbers,amsmath,amssymb]{revtex4}
\usepackage{graphicx}
\usepackage{mathptmx}
\usepackage{verbatim}
\usepackage{graphicx}
\usepackage{amsmath}
\usepackage{amssymb}
\usepackage{mathdots}
\usepackage{amsfonts}
\usepackage{mathrsfs}
\usepackage{amsmath}
\usepackage{color}
\usepackage{graphicx}
\usepackage{bm}
\usepackage{amssymb}
\usepackage{xspace}
\usepackage{dcolumn}

\begin{document}
\title{
Accurate tight-binding models for the $\pi$ bands of bilayer graphene}

\author{Jeil Jung}
\affiliation{Department of Physics, University of Texas at Austin, USA}
\affiliation{Department of Physics, National University of Singapore, Singapore}
\author{Allan H. MacDonald}
\affiliation{Department of Physics, University of Texas at Austin, USA}

\begin{abstract}  
We derive an {\em ab initio} $\pi$-band tight-binding model for $AB$ stacked 
bilayer graphene based on maximally localized Wannier wave functions 
(MLWFs) centered on the carbon sites,  finding that both intralayer and interlayer hopping is longer in range than 
assumed in commonly used phenomenological tight-binding models.   
Starting from this full tight-binding model, we derive two effective models that are intended to provide a convenient 
starting point for theories of $\pi$-band electronic properties by achieving accuracy 
over the full width of the $\pi$-bands, and especially at the Dirac points, in models with a relatively 
small number of hopping parameters.  The simplified models are then compared with 
phenomenological Slonczewski-Weiss-McClure type tight-binding models in an effort to 
clarify confusions that exists in the literature concerning tight-binding model parameter signs.  
\end{abstract}

\pacs{73.22.Pr, 71.20.Gj,71.15.Mb,31.15.aq}

\maketitle
\section{Introduction}
$\pi$-bands are responsible for the low-energy electronic properties of 
graphitic systems, including single and multilayer graphene.  It is often convenient to 
base theories of electronic properties on orthogonal orbital tight-binding models
similar to the Slonczewski-Weiss-McClure\cite{wallace,swm,tatar} 
(SWM) $\pi$-band tight-binding model used extensively in graphite.
The most important parameters in these models are the near-neighbor hopping amplitude within the  
graphene layers, which sets the $\pi$-band width, and 
the near-neighbor inter-layer hopping amplitude which partially splits\cite{mingchiral} the
two states per layer which are nearly degenerate at the Brillouin-zone corner points $K$ and $K'$.  
Although smaller in magnitude other parameters
still play a crucial role in determining electronic properties in multi-layer graphene systems, 
for example by moving level crossings (Dirac points) away from $K$ and $K'$, 
and cannot be guessed with sufficient accuracy simply by making analogies with the graphite case.
In this paper we attempt to achieve a broader perspective on these tight-binding models by 
deriving a $\pi$-band model for bilayer graphene from {\em ab initio} electronic structure calculations.  
Bilayer graphene is the simplest multi-layer system and can be viewed as 
an isolated copy of the two-layer repeating unit of Bernal or AB stacked graphite.
Tight-binding models for bilayer graphene have been proposed previously 
based on analogies to the SWM model for graphite \cite{partoens,nilsson},
on phenomenological fits to Raman \cite{ramanbilayer} and infrared \cite{infraredfit,kuzmenko} measurements,
and on parametric fits to first principles bands.\cite{abinitiofit}

The literature on tight-binding models for bilayer graphene does not 
provide consistent magnitudes or even signs for 
the small remote hopping parameters that reshape the bands near the Dirac points.
In this paper we obtain $\pi$-band tight-binding models from 
first principles LDA electronic structure calculations combined with 
maximally localized Wannier functions (MLWF) centered at the carbon sites.\cite{mlwf}
Models based on Wannier functions with well defined symmetries
benefit from the orthonormality of the localized orbitals
and give rise to matrix elements that can be physically interpreted as hopping 
between $\pi$ orbitals centered on different sites.
A $\pi$-band only model for bilayer graphene should be accurate because hybridization between $\pi$ and 
$\sigma$ orbitals is weak in all multi-layer graphene systems.\cite{saito,footnote} 
Calculations based on the LDA have the advantage that they do not 
incorporate the many-body band renormalisations due to screened non-local exchange 
that are known to be important in graphene and multi-layer graphene systems.  The resulting bands 
can therefore be used as a starting point for studies of many-body band renormalization,
as is common practice for example in GW approximations calculations.\cite{rubio}

The present work extends a previous analysis carried out for single layer graphene, \cite{graphene}
to the case of Bernal stacked bilayer graphene.
The Hamiltonian we obtain provides an intuitive 
understanding of remote hopping processes both within and 
between layers.  One important finding is that both intra-layer and inter-layer 
hopping processes have longer range than assumed in phenomenological
tight-binding models.  We therefore propose a systematic scheme which we use 
to obtain alternate effective tight-binding models with a smaller 
number of parameters which retain both accuracy near Dirac points and 
a good description of the full $\pi$-bands.
Our paper is structured as follows.
In section \ref{sec:mainsection} we summarize 
the full tight-binding (FTB) model of bilayer graphene's $\pi$-bands obtained
from a calculation that constructs maximally localized Wannier functions.
In section \ref{sec:effectivemodel} we explain the relationship between 
tight-binding and continuum models and the scheme used to derive the  
simplified tight-binding models which we hope will be a convenient starting point for 
some $\pi$-band electronic property theories.   
We then devote section \ref{swm} to a comparison between our simplified models 
and models that have been used in the literature for 
bilayer graphene systems.
Finally our conclusions are summarized in section  \ref{sec:summary}.

\begin{figure}
\includegraphics[width=7.4cm,angle=0]{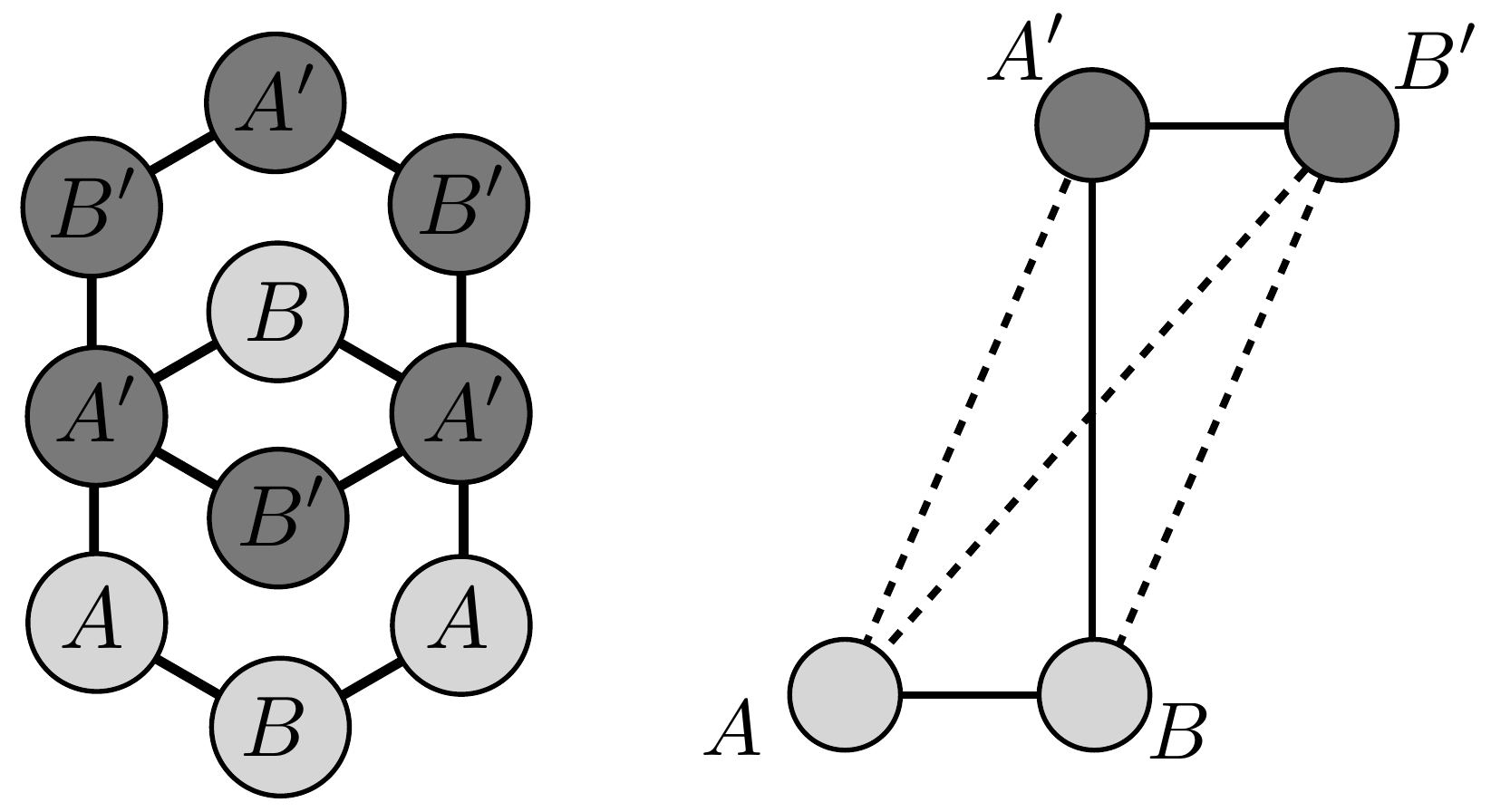}  
\caption{
Definition of the unit cell for Bernal or $AB$ stacked bilayer graphene.
The $A'$ site of the top layer sits right above a $B$ site of the lower layer.
The $A$ and $B'$ sites do not have a corresponding vertical neighbor. 
The solid lines in the right panel represent the large-amplitude
hopping processes in literature tight-binding models, whereas the dotted lines 
indicate weaker processes that are responsible for particle-hole symmetry breaking and trigonal distortions.
In the main text we quantify the role they play in bilayer graphene electronic structure.}
\label{bilayerunitcell}
\end{figure}

\section{Wannier interpolation of Bernal bilayer graphene $\pi$-bands}
\label{sec:mainsection}
\begin{figure}
\includegraphics[width=8cm,angle=0]{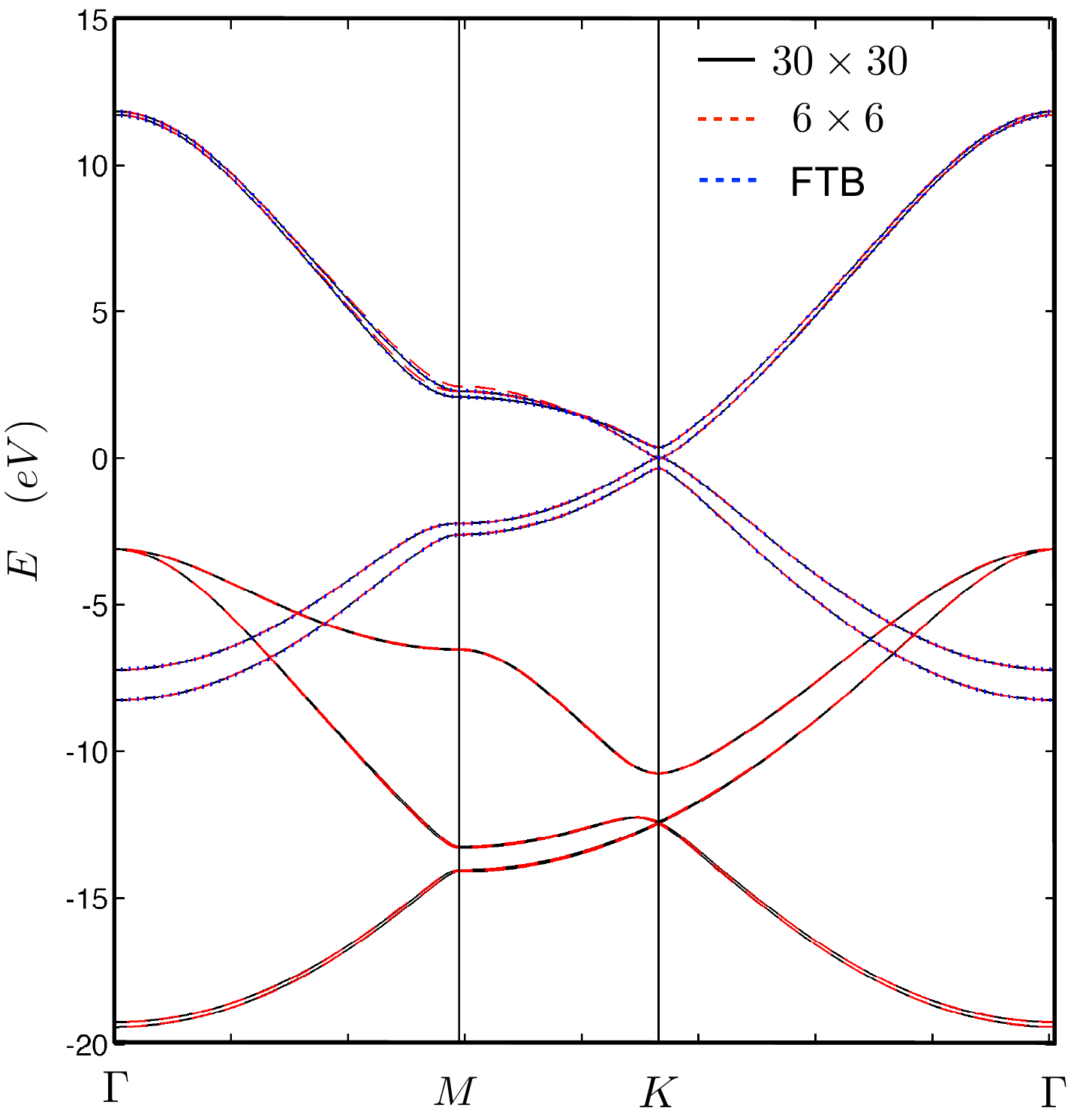} 
\caption{
(Color online) 
Band structure of $AB$ stacked bilayer graphene obtained through Wannier 
interpolation of first principles LDA results for $6 \times 6$ 
and $30 \times 30$ $k$-point sampling densities. 
The $6 \times 6$ interpolation is accurate expect near the 
$M$ point, where the maximum deviation is $\sim 0.2$ eV, 
and the $30 \times 30$ interpolation is accurate throughout the Brillouin zone.   
The difference near the $M$ point between the $6\times 6$ and $30 \times 30$ 
Wannier interpolation is larger than in the 
single-layer graphene case.
}
\label{bands}
\end{figure}

A $\pi$-band model of bilayer graphene contains four orthogonal orbitals per unit 
cell, corresponding to $p_z$ orbitals centered on the $A$, $B$ 
sites in the bottom layer and  $A^{\prime}$ and $B^{\prime}$ sites in the top layer. 
We have fixed the lattice constants at the experimental values of graphite
for which the in-plane lattice constant is $a = 2.46 \AA$ and the interlayer separation 
is $c = 3.35 \AA$.  We label the Bernal (AB) stacked honeycomb lattice sites 
so that the $A^{\prime}$ site 
on the top layer is directly above the $B$ site in the bottom layer
as illustrated in Fig. \ref{bilayerunitcell}.
Our definitions of the in-plane lattice vectors and 
Bloch functions are similar to those
used in reference [\onlinecite{graphene}] for single layer graphene.
The technical details of the calculations used to extract the Wannier functions
are also similar to those of the single layer case.
We used the plane-wave 
first principles calculation package {\em Quantum Espresso} \cite{quantumespresso}
and {\em wannier90} \cite{wannier90} to construct the Wannier functions.
The {\em Quantum Expresso} calculations used the Perdew-Zunger 
parametrization \cite{pz} of the local density approximation.\cite{lda}
The tight-binding model parameters obtained using a  
$30 \times 30$ k-space sampling density are listed in the Appendix.
Band structures resulting from Wannier interpolation for two 
different $k$-point sampling densities are presented in Fig. \ref{bands}
where we show the excellent agreement between the tight-binding model 
and the first principle calculation for $30 \times 30$ sampling,
indicating that the parameters for hopping processes more remote 
that those listed in the Appendix are entirely negligible.  

\begin{figure*}
\includegraphics[width=16.5cm,angle=0]{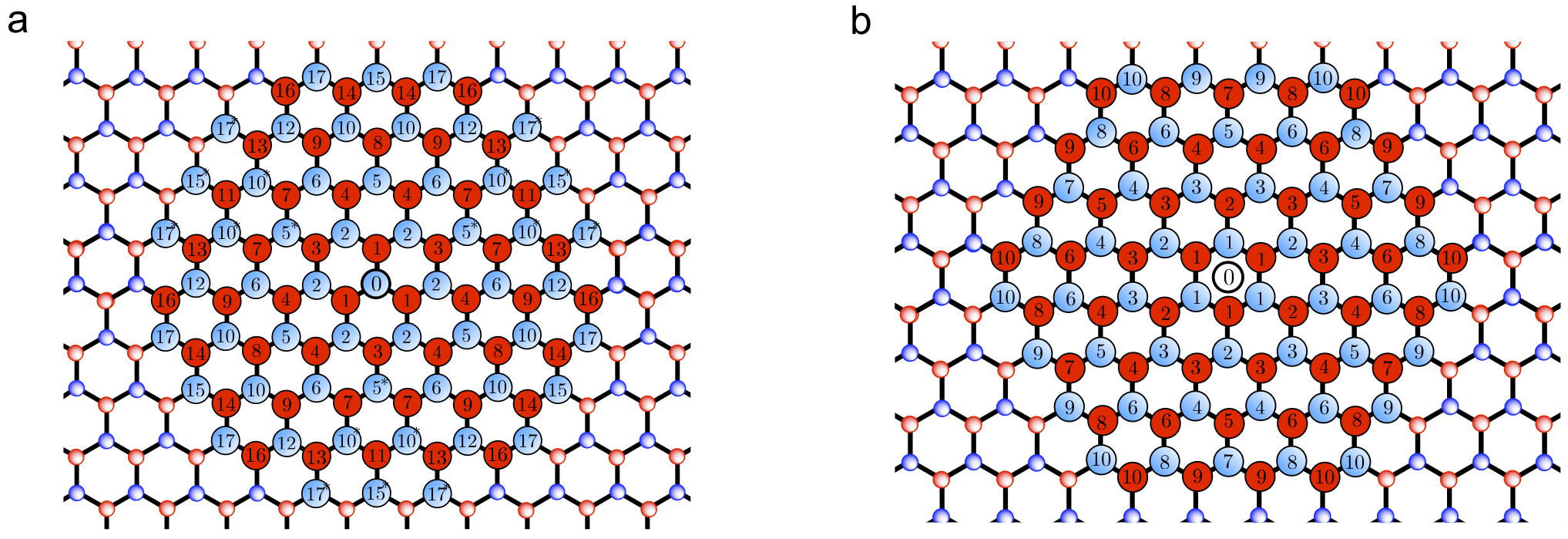}  
\caption{
(Color online)
Neighbor classification relative to a central reference point labelled $0$.
Neighbors of a given site can be grouped into symmetry equivalent 
subsets with identical hopping parameters.  The neighbor groups are ordered by their distance from the 
reference point. $m^{th}$ nearest neighbors of the reference point are labelled in this figure by the integer $m$. 
{\em Left panel:}
Neighbor classifications for intralayer $AB$ pairs and for  
interlayer $BA^{\prime}$ and $BB^{\prime}$ pairs. 
{\em Right panel:}
Neighbor classification for interlayer $AA^{\prime}$, $AB^{\prime}$ pairs.
The structure factors (See text) in $AA^{\prime}$ matrix elements are
identical to those in $AB$ elements.  
The structure factors in $AB^{\prime}$ acquires a mirror reflection in the $y$ direction
that leads to a complex conjugation in the matrix elements.
The integer labels are derived by grouping neighbors according to in-plane separation 
so that some $AA^{\prime}$ and $AB^{\prime}$ neighbor pairs share labels
with pairs in the left panel.
}
\label{cutoffs}
\end{figure*}

\begin{figure}
\includegraphics[width=7.6cm,angle=0]{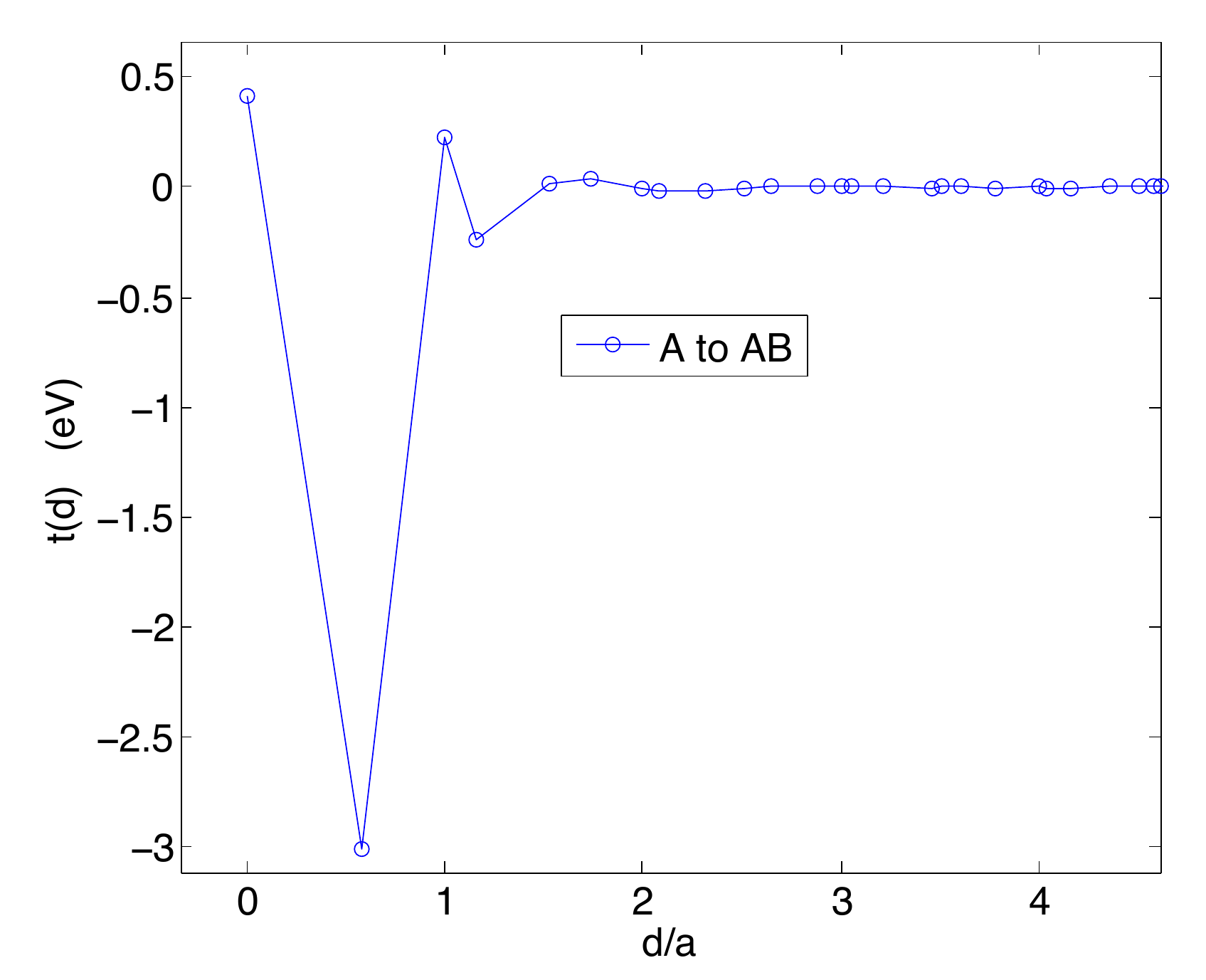}  
\caption{
(Color online)
Intralayer hopping terms in Bernal bilayer graphene evaluated 
starting from $A$ to $A$ and $B$ sites in the bottom layer  
with $30 \times 30$ $k$-point sampling density.
The intralayer Hamiltonian matrix elements are extremely similar to those
of isolated graphene layers obtained in Ref. [\onlinecite{graphene}] 
in spite of the interlayer coupling effect.
The difference between isolated layer and bilayer intralayer hopping 
is not visible on the scale of this figure.
}
\label{intraa}
\end{figure}

The $\pi$-band Hamiltonian for bilayer graphene is 
a ${\bf k}$-dependent four dimensional matrix:
\begin{eqnarray}
{H} ({\bf k}) 
&=&  \begin{pmatrix}
  H_{B} ({\bf k})                         &      H_{BT} ({\bf k})         \\  
  H_{TB} ({\bf k})                       &      H_{T} ({\bf k})        
\end{pmatrix}
\label{hamiltonian1} 
\end{eqnarray}
where $H_{B}({\bf k})$ and $H_{T}({\bf k})$ are two dimensional matrices
that describe intra and inter sub lattice hopping within the bottom and top graphene layers.
The Hamiltonian of the bottom graphene sheet is 
\begin{eqnarray}
{H}_{B} ({\bf k}) 
&=&  \begin{pmatrix}
  H_{AA} ({\bf k})                       &      H_{AB} ({\bf k})          \\  
  H_{BA} ({\bf k})                       &      H_{BB} ({\bf k})          
\end{pmatrix}, 
\end{eqnarray}
and the Hamiltonian of the top graphene sheet 
\begin{eqnarray}
{H}_{T} ({\bf k}) 
&=&  \begin{pmatrix}
  H_{A^{\prime}A^{\prime}} ({\bf k})                       &      H_{A^{\prime}B^{\prime}} ({\bf k})          \\  
  H_{B^{\prime}A^{\prime}} ({\bf k})                       &      H_{B^{\prime}B^{\prime}} ({\bf k})          
\end{pmatrix}. 
\end{eqnarray}
Both intralayer and interlayer hopping processes can be classified into symmetry equivalent groups.
The grouping of intralayer processes is identical to that in single-layer graphene.\cite{graphene}
The intralayer Hamiltonian can be written as a sum over groups of the product of a hopping strength for the group 
and a structure factor: 
\begin{eqnarray}
H_{A(')B(')} ({\bf k}) =   \sum_{n} t_{A(') B(') n} \,\, f_{n}({\bf k}) \\
H_{B(')A(')} ({\bf k}) =   \sum_{n} t_{B(') A(') n} \,\, f_{n}^*({\bf k}) \\
H_{A(')A(')} ({\bf k}) =   \sum_{n} t^{\prime}_{A(') A(') n} \,\, g_{n}({\bf k})  \\
H_{B(')B(')} ({\bf k}) =   \sum_{n} t^{\prime}_{B(') B(') n} \,\, g_{n}({\bf k}).
\end{eqnarray}
The structure factors $f_{n}({\bf k})$ and $g_{n}({\bf k})$ are defined in Ref. [\onlinecite{graphene}].  
Inversion symmetry leads to the following relations between the hopping amplitudes 
in top and bottom layers: 
\begin{eqnarray}
t_{A B n} &=& t_{B^{\prime} A^{\prime} n}, \quad t^{\prime}_{A A n} 
= t^{\prime}_{B^{\prime} B^{\prime} n}, \quad 
t^{\prime}_{B B n} = t^{\prime}_{A^{\prime} A^{\prime} n}.
\end{eqnarray}
The  intralayer hopping amplitude values are extremely close to those obtained for 
isolated graphene layers as summarized in Fig. \ref{intraa}.  

\begin{figure}[htbp]
\includegraphics[width=8cm,angle=0]{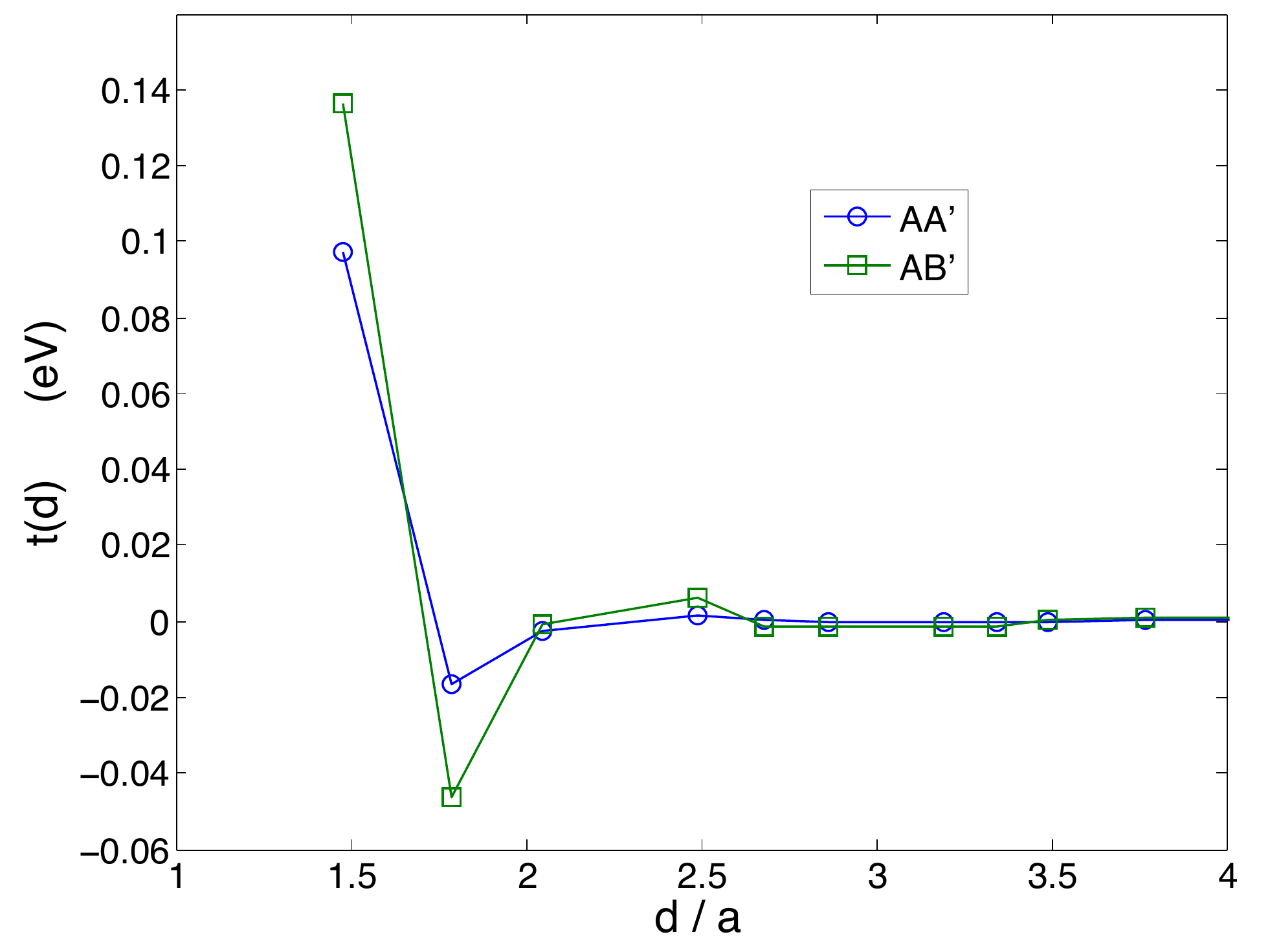}  \\
\includegraphics[width=7.8cm,angle=0]{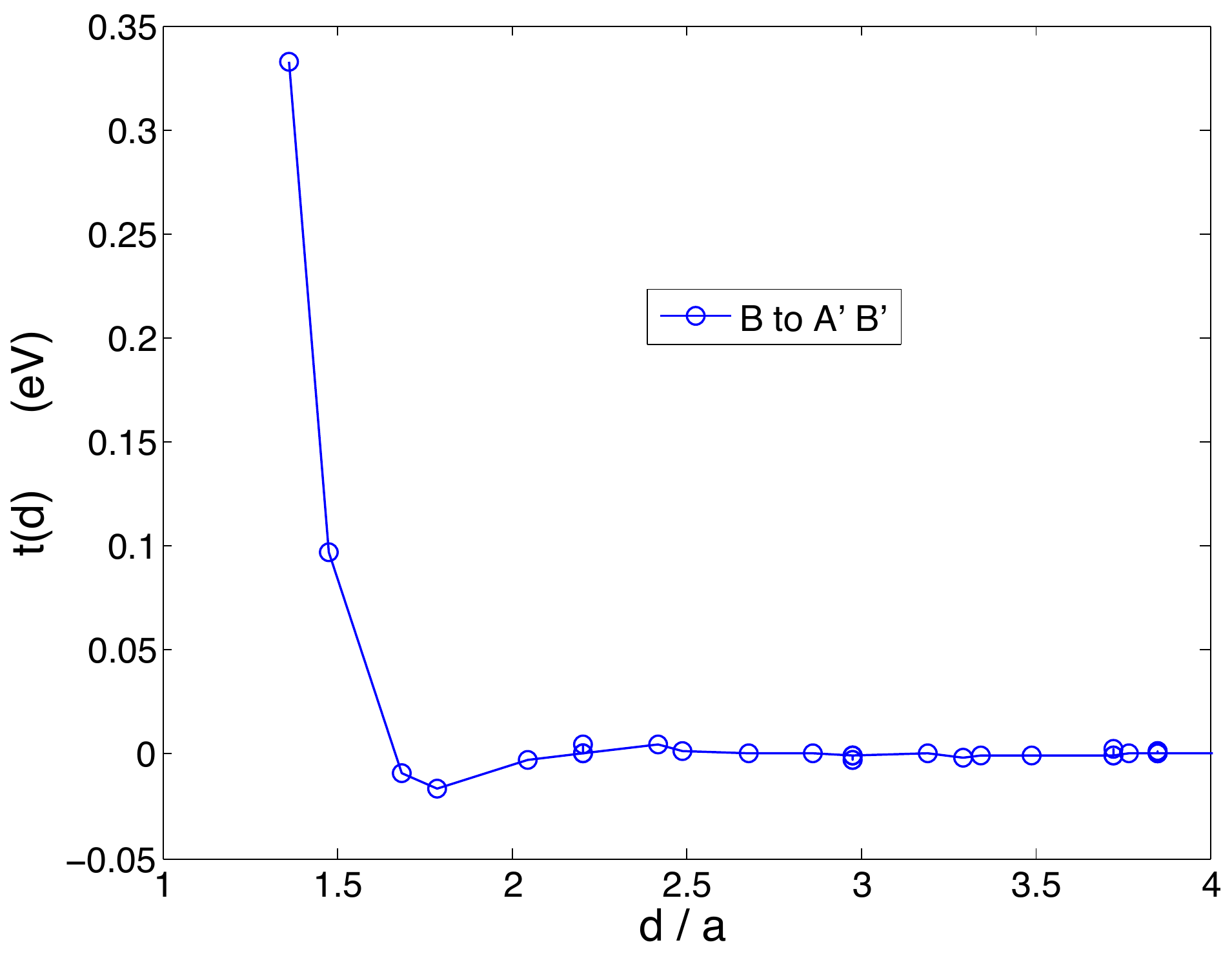}  
\caption{
(Color online)
{\em Upper panel:}
Hopping parameters connecting site $A$ of the bottom layer to the $A^{\prime}$
and $B{^{\prime}}$ sites in the top layer. 
The distances on the horizontal axis correspond to 
in-plane projections of displacements.
{\em Lower panel:}
Hopping parameters connecting site $B$ of the bottom layer to the top layer sites.
Note that two different values occur for some values of $d$ ($m=5, 10, 15, 17$) 
because of the occurrence of 
neighbor pairs which share the same projected displacements that are not 
symmetry equivalent.  These cases are marked with an $*$ symbol
in Fig. \ref{cutoffs}.
}
\label{interlayercouplingAB}
\end{figure}

Coupling between layers is described by the  
$H_{BT}({\bf k}) = H^{*}_{TB}({\bf k})$ where
\begin{eqnarray}
{H}_{BT} ({\bf k}) 
&=&  \begin{pmatrix}
  H_{AA^{\prime}} ({\bf k})                       &      H_{AB^{\prime}} ({\bf k})          \\  
  H_{BA^{\prime}} ({\bf k})                       &      H_{BB^{\prime}} ({\bf k})          
\end{pmatrix}.
\end{eqnarray}
We find that 
\begin{eqnarray}
H_{AA^{\prime}} ({\bf k}) =   \sum_{n} t_{A A^{\prime} n} \,\, f_{ n}({\bf k})  \\
H_{AB^{\prime}} ({\bf k}) =   \sum_{n} t_{A B^{\prime} n} \,\, f^{*}_{ n}({\bf k})  \\
H_{BA^{\prime}} ({\bf k}) =   \sum_{n} t^{\prime}_{B A^{\prime} n} \,\, g_{ n}({\bf k})  \\
H_{BB^{\prime}} ({\bf k}) =   \sum_{n} t_{B B^{\prime} n} \,\, f_{ n}({\bf k}).   \label{hamiltonianlast}
\end{eqnarray}
where the interlayer structure factors are related to intralayer structure factors by 
$f_n = f_{ABn} = f_{A^{\prime} B^{\prime}n} = f_{AA^{\prime}n} = f^{*}_{AB^{\prime}}$
and $g_n = g_{AA n} = g_{BBn} = g_{A^{\prime}A^{\prime}n} 
= g_{B^{\prime}B^{\prime}n} = g_{BA^{\prime}n}$, as is 
apparent when we compare the lattice sites illustrated in Fig. \ref{cutoffs}.
The interlayer hopping terms are plotted in Fig. \ref{interlayercouplingAB}
as a function of the in-plane projection of the hopping distances.
Two hopping terms are prominent but, as we will discuss later on, smaller terms do 
play a role in defining details of the electronic structure
both near and far away from the Dirac point .

\section{Simplified effective models with fewer parameters}
\label{sec:effectivemodel}

Our maximally-locallized Wannier function calculations demonstrate that
the number of important interlayer hopping processes in few layer graphene systems,
and by implication also in graphite, is larger than in SWM-inspired 
phenomenological models.  These phenomenologies provide a good description
of states near the Dirac point, but should not be taken literally as a statement concerning
the relative strengths of different microscopic processes on an 
atomic length scale.  In the following we discuss the construction of 
models that are partially in the SWM spirit, in the sense that they 
concentrate on accuracy near the Dirac point, but are informed by the 
full maximally localized Wannier function calculation.  For this purpose we first
carefully discuss the Dirac-point low-energy continuum model implied by 
particular tight binding models.   

\subsection{Continuum model near the Dirac point}
\label{continuum}
We write the four dimensional $\pi$-band Hamiltonian in the form  
\begin{eqnarray}
{H} ({\bf k})     &=&  
\begin{pmatrix}
      G_{AA} ({\bf k})                      &      F_{AB} ({\bf k})      
  &  F_{AA^{\prime}} ({\bf k})       &      F^*_{AB^{\prime}} ({\bf k})    \\  
  &  G_{BB} ({\bf k})                     &      G_{BA^{\prime}} ({\bf k})                      
  &  [F_{AA^{\prime}} ({\bf k})]    \\ 
  \vdots           &         \ddots 
  &  [G_{BB} ({\bf k})]    & [F_{AB} ({\bf k})]    \\ 
  &          \hdots     
  &    &  [G_{AA} ({\bf k})]     
\end{pmatrix},
\label{hamiltonian}
\end{eqnarray}
where we have labeled the Hamiltonian matrix elements by the letters
$F$ and $G$ to emphasize that they consist of sums of the $f_n$ and $g_n$
structure factors respectively.  
The matrix elements surrounded by square brackets are equivalent to other matrix elements by 
the symmetry relations $F_{BB'}({\bf k}) = F_{AA'}({\bf k}) $, $F_{A'B'}({\bf k}) = F_{AB}({\bf k}) $, 
$G_{A'A'}({\bf k}) = G_{BB}({\bf k}) $, and $G_{B'B'}({\bf k}) = G_{AA}({\bf k}) $.
We discuss all the inequivalent Hamiltonian matrix elements below.  
Note that $F_{\alpha \beta}({\bf k}_D) = 0$, that the intralayer non-zero values of
$G_{\alpha \beta}({\bf k}_D)$ can be interpreted as site energy shifts, and 
that the interlayer non-zero values of $G_{\alpha \beta}({\bf k}_D)$  can be interpreted as 
inter-layer tunneling amplitudes.  The non-zero values of $G_{\alpha \beta}({\bf k}_D)$ play the 
most essential role in defining the low energy physics of bilayers.
We choose our zero of energy so that $G_{AA}({\bf k}_D) = G_{B'B'}({\bf k}_D) = 0$.

%
\begin{figure}
\includegraphics[width=8.5cm,angle=0]{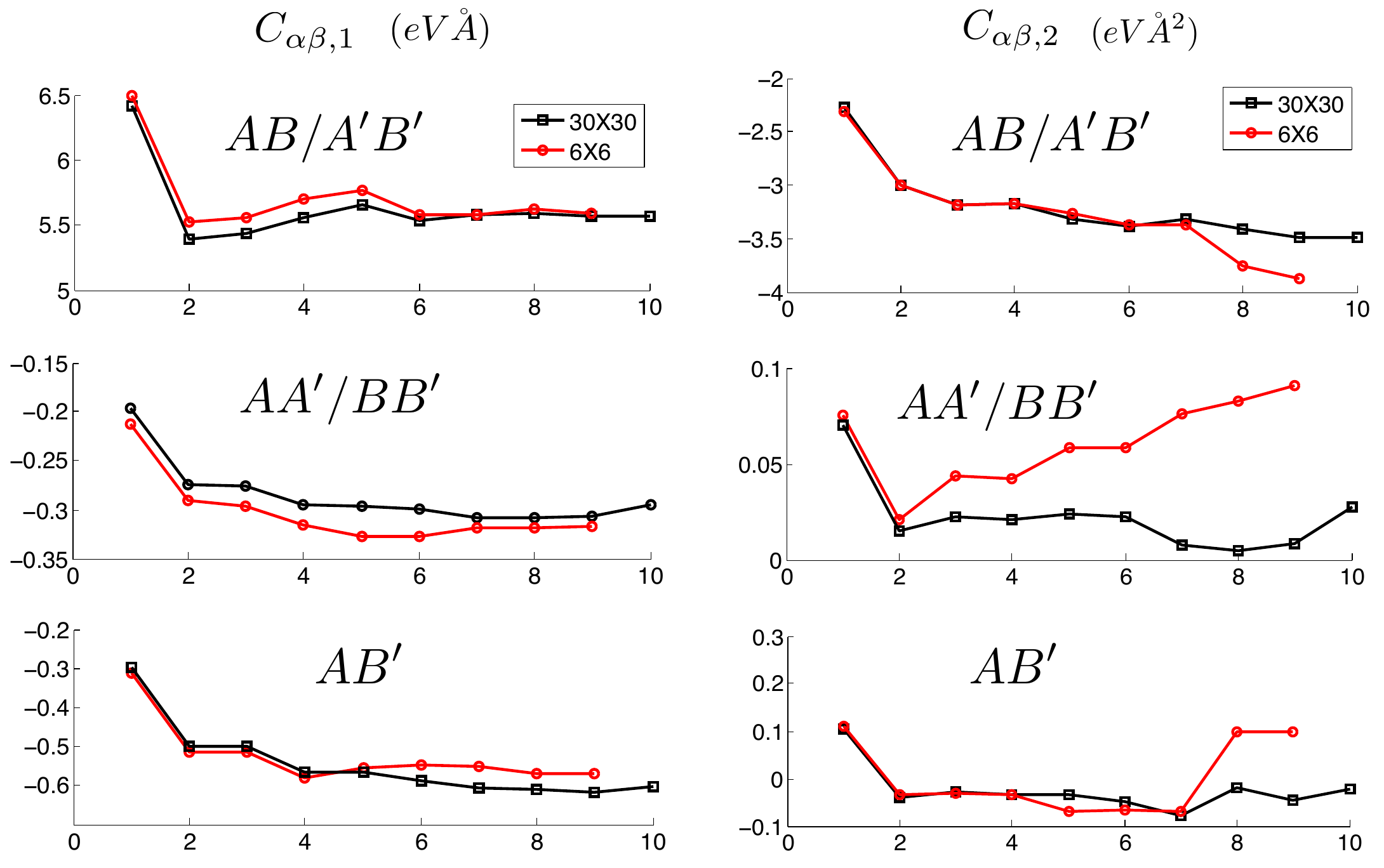} 
\caption{
(Color online)  
Evolution of the $\vec{k} \cdot \vec{p}$ model 
coefficients $C_{\alpha \beta 1}$ and $C_{\alpha \beta 2}$
with neighbor number truncation.   
These coefficients are used in construction of the 
effective tight-binding model in Eqs. (\ref{simpletb1}-\ref{simpletb3}).
The comparison of results obtained by $6\times6$ and $30\times30$ calculations
points to the need for denser sampling to capture 
distortions of the local $p_z$ orbitals by interlayer coupling.
}
\label{comp}
\end{figure}

The matrix elements can be expanded up to quadratic order using 
Eqs. (11-16) of Ref. [\onlinecite{graphene}].
For $\alpha \beta = AB, AA^{\prime}, B A^{\prime}$, 
and wave vectors near ${\bf k}_D = (4\pi/3a,0)$ we write 
\begin{eqnarray}
F_{\alpha \beta}({\bf k}_D + {\bf q}) &\simeq&  
    C_{\alpha \beta 1} q e^{-i \theta_{\bf q}} 
+  C_{\alpha \beta 2} q^2 e^{i 2 \theta_{\bf q}}.
\label{ffunc}
\end{eqnarray}
The expansion of the $F_{\alpha \beta}$ functions near $(-4\pi/3a,0)$ differs by 
an overall sign and complex conjugation.
The dependence of these $\vec{k} \cdot \vec{p}$ Hamiltonian model 
coefficients on truncation of the full tight binding model 
is illustrated in Fig. \ref{comp}.
For $\alpha \beta = AA, BB, BA^{\prime}$ we write 
\begin{eqnarray}
G_{\alpha \beta}({\bf k}_D + {\bf q}) &\simeq& C^{\prime}_{\alpha \beta 0} 
+ C^{\prime}_{\alpha \beta 2} \, q^2  
\label{gfunc}
\end{eqnarray}
When the small $q$ expansions
in Eqs. (\ref{ffunc}-\ref{gfunc}) are used for the matrix elements of  Eq. (\ref{hamiltonian})  
the electronic structure near the Dirac point is accurately reproduced.

\begin{figure}
\includegraphics[width=8cm,angle=0]{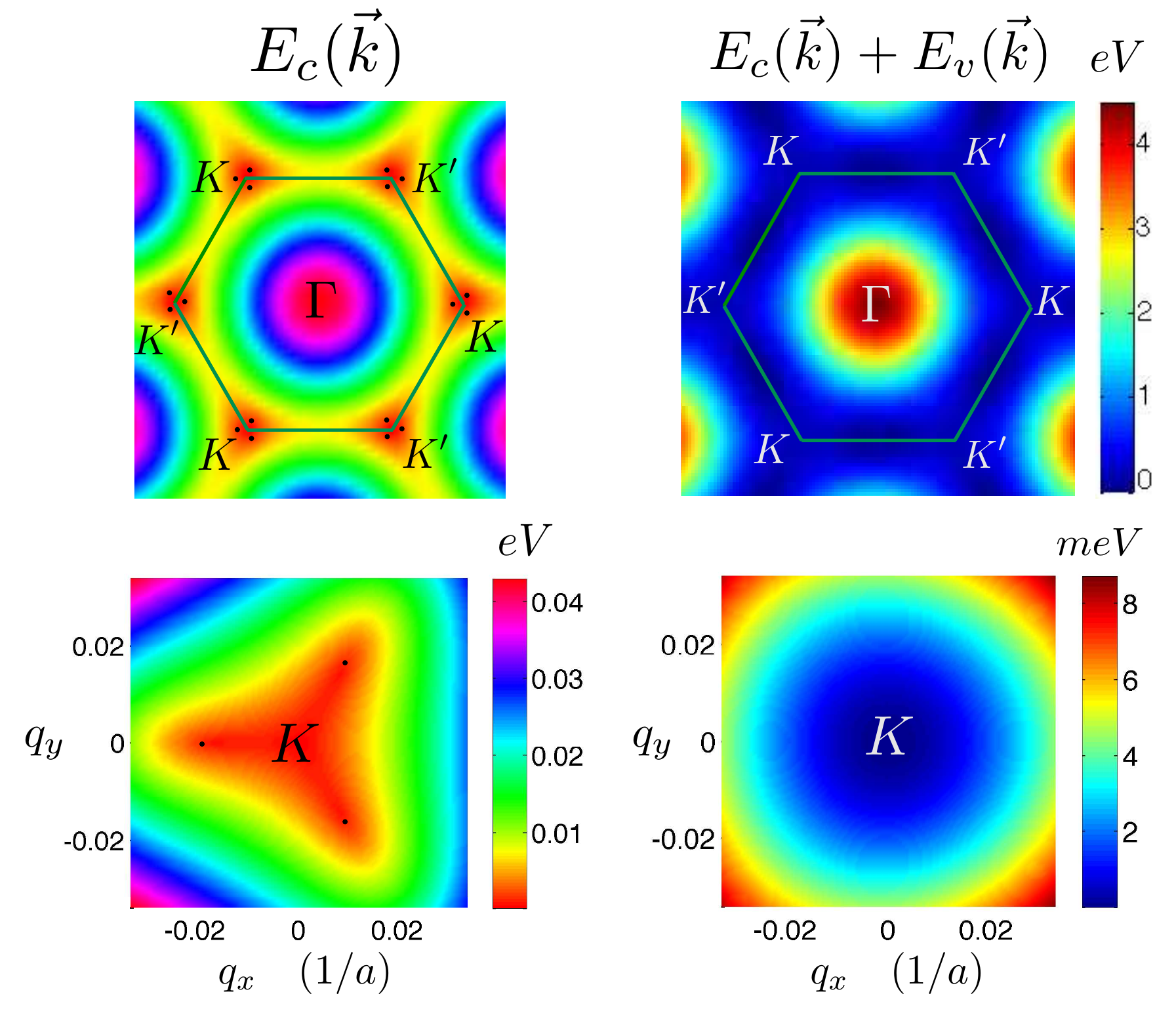} 
\caption{
(Color online)  
Two dimensional color scale plot of the LUMO band dispersion  
over the full Brillouin zone (top left) and near the ${\bf k}_D = (4\pi/3a,0)$
Dirac point.  The points at which the conduction and valence bands 
are degenerate are represented by black dots which lie  
at the vertices of a triangle.
{\em Right panel:}
Particle-hole symmetry breaking illustrated by plotting the sum of
the LUMO (lowest conduction) and HOMO (highest valence) band energies ($E_c({\bf k}) + E_{v}({\bf k})$).
The main contribution comes from the $AA'$ and $BB'$ hopping processes
as explained in the main text.
}
\label{trigph}
\end{figure}

\begin{figure}[h]
\includegraphics[width=8cm,angle=0]{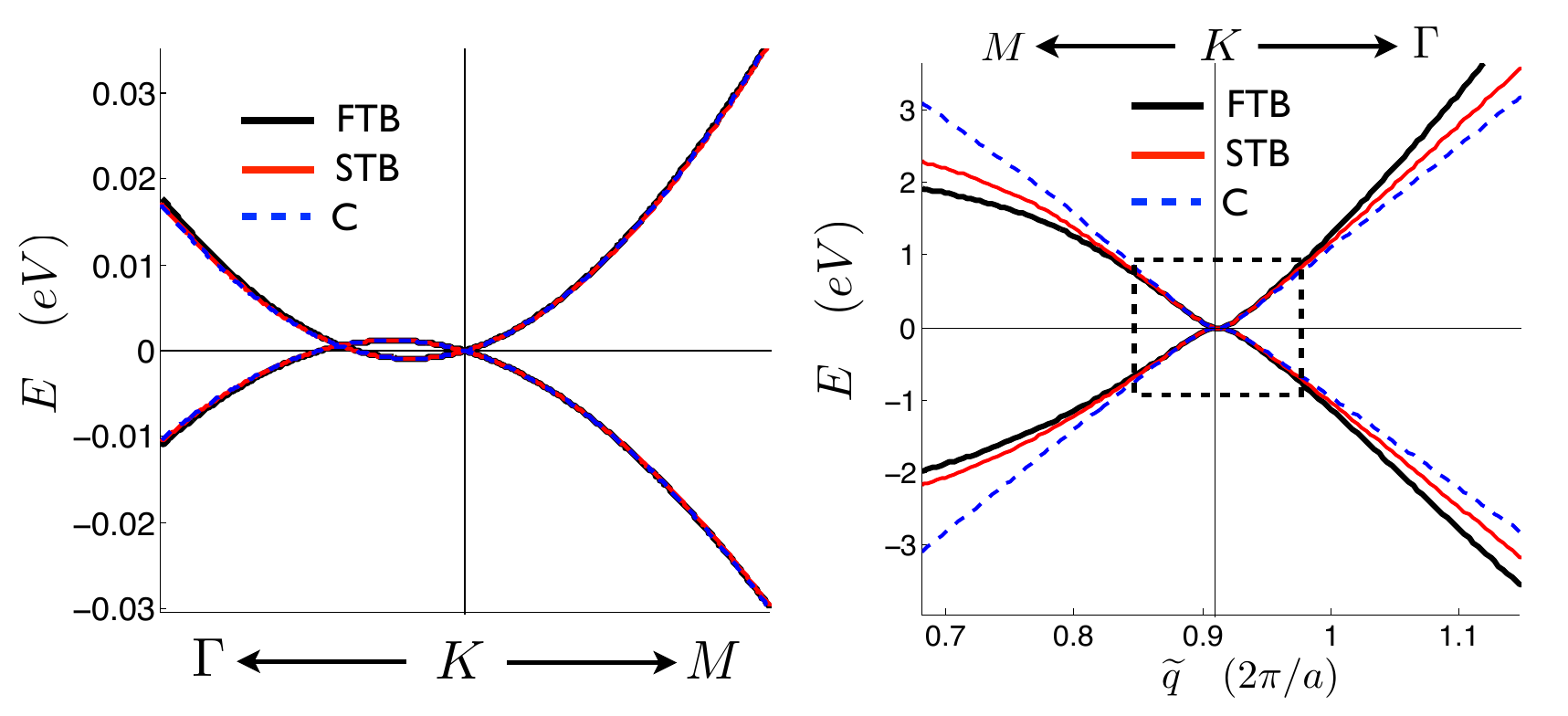} 
\caption{
(Color online)  
Comparison of electronic structure near the Dirac point
for the full tight-binding model (FTB) model, 
the simplified tight-binding model explained in the text (F1G0),
and the effective continuum model (C) defined by Eqs. (\ref{hamiltonian}-\ref{gfunc}) 
with the second order term in $k$ neglected.  
Deviations increase as we move away from the Dirac point, but the 
relative error is of the order of a few \% for $q \sim 1/a$ and F1G0 is 
superior to the continuum approximation. 
The dotted square represents the region
over which the deviation from the Dirac point $\left| q \right| < 1/a$.
}
\label{compare}
\end{figure}

\subsection{Single structure factor tight-binding model}
\label{singlesf}
A simplified model that is reminiscent of the 
Slonczewski-Weiss and McClure (SWM) model for graphite \cite{swm}
can be constructed by identifying the shortest range hopping
model that can capture the correct zeroth order terms in the expansion of 
$G_{\alpha \beta}({\bf k}_D + {\bf q} )$ and the correct 
first order terms in the expansion of $F_{\alpha \beta}({\bf k}_D + {\bf q} )$.
Given a choice for the zero of energy this leaves in a model with 
five independent parameters:
\begin{eqnarray}
H_{AA} ({\bf k}) &=&  H_{BB'} ({\bf k}) =  t'_{AA 0}    \label{simpletbp1} \\
H_{BB} ({\bf k}) &=&   H_{A' A'} ({\bf k}) = t'_{BB 0}   \label{simpletbp2} \\
H_{BA'} ({\bf k}) &=&    t'_{BA'0}.  \label{simpletbp3}
\end{eqnarray}
and 
\begin{eqnarray}
H_{A B} ({\bf k}) &=&  H_{A'B'} ({\bf k}) =   t_{AB 1}  \, f_1({\bf k})    \label{simpletb1}   \\
H_{A A'} ({\bf k}) &=&  H_{BB'} ({\bf k}) =    t_{AA' 1}    \, f_1({\bf k})   \label{simpletb2}   \\
H_{A B'} ({\bf k}) &=&      t_{AB'1}   \, f^*_1({\bf k}).       \label{simpletb3}
\end{eqnarray}
Here $f_1 ({\bf k}) = \exp( i  k_y a/\sqrt{3} )
+ 2 \exp(  -i k_y a/ 2 \sqrt{3}   ) \cos( k_x a / 2 )$ with the correct linear dispersion coefficient: 
with $ t_{\alpha \beta 1} = - 2 C_{\alpha \beta,1} / {\sqrt{3} a }$.
The effective tight-binding parameters in this model, listed in Table \ref{table:effective}, 
are not the same as the physical hopping parameters listed in the Appendix.
This approximation folds all hopping parameters down to a model with 
only near-neighbor and on-site or purely vertical type hopping, but 
has the same continuum model limit as the full tight-binding model.  
This model achieves accuracy near the Dirac point, and is more 
accurate across the full Brillouin-zone than the continuum model as illustrated in Fig. \ref{compare}.

\begin{table}
  \begin{tabular}{|  c |  c c c |  c |   c c |}   \hline
\multicolumn{1}{|c|}{ }         &
\multicolumn{3}{c|}{ $F_{\alpha \beta}$}         &
\multicolumn{1}{c}{} &
\multicolumn{2}{c|}{ $G_{\alpha \beta}$  }     
  \\    \hline
  \multicolumn{1}{|c|}{$\alpha \beta$}                        &
    \multicolumn{1}{c}{ $t_{\alpha \beta,1}$}             &
  \multicolumn{1}{c}{ $C_{\alpha \beta 1}$}                &
  \multicolumn{1}{c|}{ $C_{\alpha \beta 2}$}             &
  \multicolumn{1}{|c|}{ $\alpha \beta$}         &
\multicolumn{1}{c}{ $  
t'_{\alpha \beta,0}$  }     &
\multicolumn{1}{c|}{$C_{\alpha \beta 2}^{\prime}$ }     
\\  \hline
  $AB$, $A' B'$                  &       $-2.61$    &  5.567        &  $ -3.494    $     &   $AA$         &    
 0   &  $-0.269$    \\
  $A A^{\prime}$, $BB'$     &    $0.138 $ &    $-0.2949   $   &  0.0272          &   $BB$            &     
  0.015 & $-0.222$    \\  
  $AB^{\prime}$                 &   $ 0.283 $  &   $ -0.6036 $    &  $-0.0227$      &   $B A^{\prime}$        &     
   0.361 &  0.103   \\
 \hline
  \end{tabular}
  \quad
\caption{
$\vec{k} \cdot \vec{p}$ expansion coefficients of the  
full tight-binding model and parameters of the short-range 
tight-binding model which reproduces $C'_{\alpha \beta 0}$, $C_{\alpha \beta 1}$
and $C_{\alpha \beta 2}^{(\prime)}$. (see text) 
Note that $C_{\alpha \beta 0}' = t'_{\alpha \beta 0}$ and  $t_{\alpha \beta 1} = -2 C_{\alpha \beta 1} / \sqrt{3}a$.
The units of $t_{\alpha \beta 1}$ and $t'_{\alpha \beta 0}$ are in eV 
and the units of $C_{\alpha \beta 1}$ and $C_{\alpha \beta 2}^{(\prime)}$ are in eV$\, \AA$
and  eV$\, \AA^2$ respectively.  
}
\label{table:effective}
\end{table}

\subsection{Higher order structure factor models}
\label{twosfs}
\begin{figure}
\includegraphics[width=8cm,angle=0]{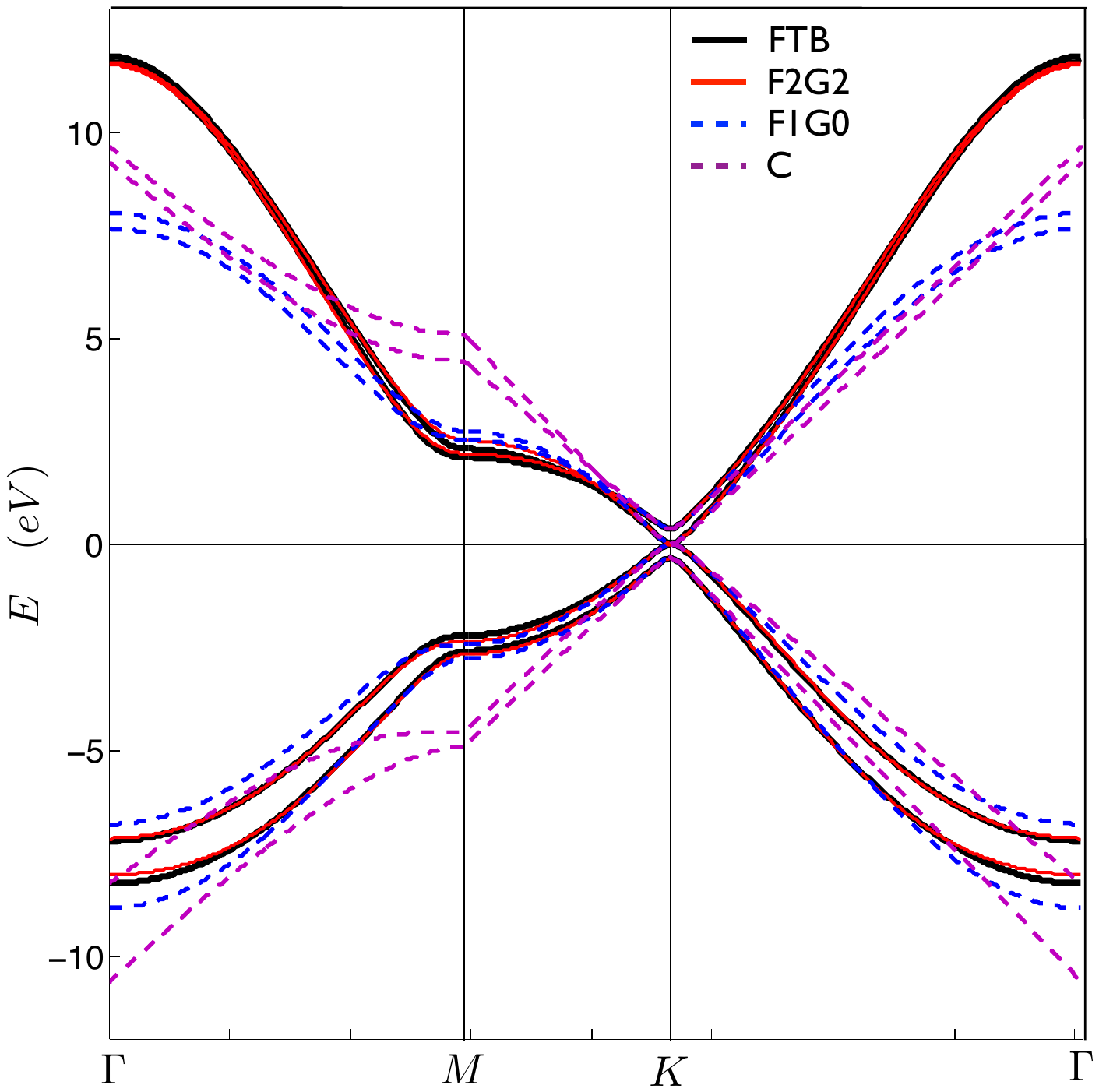} 
\caption{
(Color online) 
Comparison of the full tight-binding bands (FTB) as presented in section \ref{sec:mainsection}, 
and the simplified models presented in section \ref{sec:effectivemodel}
consisting of the continuum model (C),
the simplified tight-binding models with a single $f_1$ structure factor (F1G0),
and with structure factors up to $f_2$ and $g_2$  (F2G2).
We can observe that the simplest models C, F1G0 have substantial deviations
away from the Dirac point whereas F2G2 provides a good compromise between
simplicity and accuracy.
}
\label{effbands}
\end{figure}

The accuracy of effective tight-binding models 
far from the Dirac point can be improved by increasing the number 
of hopping parameters and associated structure factors.
In Ref. [\onlinecite{graphene}] for single layer graphene we saw that a five parameter model
with two inter-sublattice $f_n$ type structure factors
and three intra-sublattice
$g_n$ type structure factors is able to capture both the trigonal 
distortion of the bands near the Dirac points and particle-hole symmetry
breaking throughout the Brillouin zone.  One useful recipe to systematically increase the number of parameters 
in the effective model is to increase the number of structure factors $n$ 
(and hence the range of the effective tight-binding model) 
while maintaining correct values for the zeroth and first order
$\vec{k} \cdot \vec{p}$ expansion coefficients and also correct values for the
strongest shortest range hopping parameters which have 
a dominant influence on the $\pi$-band width.  
The $n=2$ truncation in the expansions of both the $F$ and $G$ functions leads to what 
we refer to as the F2G2 model where we use the correct near-neighbor hopping terms in the Appendix
for the shortest hops and correct the more distant $n=2$ hopping amplitudes using the relations
\begin{eqnarray}
t_{\alpha \beta 2} &=& C_{\alpha \beta 1}/\sqrt{3}a + t_{\alpha \beta 1} /2 \\
t'_{\alpha \beta 2} &=& \frac{1}{6} \left( C'_{\alpha \beta 0} - t'_{\alpha \beta 0} + 3 t'_{\alpha \beta 1} \right)
\end{eqnarray}
to recover 
the correct $\vec{k} \cdot \vec{p}$ expansion coefficients.  
This leads to a fifteen hopping parameter model whose parameters are listed in Table \ref{table:renormalised}.
The Hamiltonian matrix can be constructed from these parameters using Eqs. (\ref{hamiltonian1}-\ref{hamiltonianlast}).
A similar procedure can be applied when we truncate at larger values of $n$.
\begin{table}[ht] 
\centering 
\begin{ruledtabular}
\begin{tabular}{|  r   | r r r   | } 
\multicolumn{1}{|r|}{ }        &
\multicolumn{1}{r} { $t_{A B n}$ }        & 
\multicolumn{1}{r} { $t_{A A^{\prime} n}$ }       &   
\multicolumn{1}{r|} { $t_{A B^{\prime} n}$ }       
  \\   \hline
 $f_1$   &     $-3.010$   &  0.09244    &   0.1391   \\
 $f_2$   &     $-0.1984$   &   $-0.02299$      &  $ -0.07211$    \\  
\end{tabular}
\begin{tabular}{|  r   | r r r   | } 
  \multicolumn{1}{|r|}{ }        &
\multicolumn{1}{r} { $t_{A A n}^{\prime}$ }     &
\multicolumn{1}{r} { $t_{B B n}^{\prime}$ }       &
\multicolumn{1}{r|} { $t_{B A^{\prime} n}^{\prime}$ }       
  \\   \hline
  $g_0$ &    0.4295    &  0.4506   &   0.3310      \\
  $g_1$ &    0.2235    &  0.2260   &   $-0.01016$    \\  
  $g_2$ &    0.04016    &  0.0404   &   0.0001    \\
 \end{tabular}
 \end{ruledtabular}
 \caption{ 
 Effective hopping parameters in eV units for the F2G2 tight-binding model.
(See text.)  
The $n=0$ and $1$ parameters associated with the
$f_1$, $g_0$ and $g_1$ structure
factors use the microscopic hopping amplitudes whereas
the hopping terms corresponding to $f_2$ and $g_2$ have been adjusted  
to recover correct values for the  
$\vec{k} \cdot \vec{p}$ expansion coefficients $C_{1\alpha \beta}$ and $C'_{0 \alpha \beta}$. 
}
\label{table:renormalised}
\end{table}
The overall improvement in the quality of the band structure in the entire Brillouin zone
is clearly shown in Fig. \ref{effbands}.

\section{Tight-binding hopping and SWM graphite model parameters}
\label{swm}
It is instructive to compare the tight-binding models 
discussed in the previous section with bilayer graphene and graphite models
used in the literature.  Comparison with the SWM model of 
graphite, \cite{swm} which has often been used as a guide for 
band structure parameters in AB stacked few layer graphene systems including Bernal bilayer graphene,
are particularly relevant.  
The SWM model has been popular because of the attractive simplicity of having a
reduced set of 7 parameters which define the $\pi$-bands.   
Partoens and Peeters\cite{partoens} have made 
a comparison between tight-binding and SWM model band energies of graphite along two high symmetry lines 
to obtain equivalence relationships between the parameters of the 
two models.  We use  $\gamma_i$ to designate the SWM parameters. 
In Appendix B we repeat the analysis in Ref. [\onlinecite{partoens}] and find that  
\begin{eqnarray}
t_0 &\equiv& t_{AB 1} = - \gamma_0, \quad  t_1 \equiv  t'_{BA' 0} =  \gamma_1,   \nonumber \\
t_2 &\equiv & t'_{AA'' 0} =  \gamma_2/2, \quad t_3 \equiv t_{AB' 1}  =  \gamma_3,   \nonumber   \\
t_4 &\equiv&  t_{AA' 1} =  \gamma_4,  \quad t_5 \equiv  t'_{BB'' 0} = \gamma_5/2, \nonumber  \\
\Delta' &\equiv & t'_{BB 0} = \Delta - \gamma_2 + \gamma_5.
\end{eqnarray}
We have defined a new notation for our tight-binding hopping parameters above
to facilitate comparison with the SWM model parameters.
(Note that the notation $\Delta = \gamma_6$ is used in some papers.)
Even though $\gamma_2 = \gamma_5 = 0$ in bilayer graphene we considered a fictitious
third layer with ABA stacking order whose sites are labeled with $A''$ and $B''$ to complete
the equivalence relations between the SWM and the simplified tight-binding model.
The tight-binding parameters used in the comparison with the SWM model are 
the same as those used for the 
single structure factor F1G0 Hamiltonian in Eqs. (\ref{simpletbp1}-\ref{simpletb3}).
We note that the nearest neighbor interlayer hopping 
parameters in the tight-binding and SWM model  are defined so that they differ by a sign.  
The positive sign normally used for $\gamma_0$ 
is consistent with our finding that the 
numerical value of $t_{0}$ is negative.

\begin{table*}
\begin{tabular}{|   c | c |c  | c |   c | c | c | c | c |   }   \hline  
TB                    &  SWM here            &   SWM Ref. [\onlinecite{partoens}] &  SWM Ref. [\onlinecite{mccannrev}]   &  Present (LDA)    &   
Graphite [\onlinecite{charlier}]      & Bilayer [\onlinecite{ramanbilayer}]  &  Bilayer [\onlinecite{infraredfit}]  &  Bilayer [\onlinecite{kuzmenko}]  \\ \hline
$t_0 = t_{AB,1}$  & $- \gamma_0$    &  $\gamma_0$  &  $- \gamma_0$  & $-2.61$    &     $-2.598$           & $-2.9$ & $-3.0$ &   $-3.16$ \\  
$t_1 = t'_{BA',0}$  & $ \gamma_1$     &  $\gamma_1$ &  $ \gamma_1$  & 0.361     &        0.377         & 0.30 &  0.40 &   0.381   \\  
$t_2 = t'_{AA'',0}$  & $ \gamma_2/2$  &  $\gamma_2/2$ & --- &  ---         &     $-0.007$   &  --- &   ---  & ---  \\ 
$t_3 = t_{AB',1}$  & $ \gamma_3$     & $\gamma_3$  &  $ -\gamma_3$  & 0.283     &    0. 319              & 0.10 &  0.3  &  0.38  \\ 
$t_4 = t_{AA',1}$  & $ \gamma_4$     &  $-\gamma_4$  &  $ \gamma_4$ & 0.138     &    0.177        & 0.12 &  0.15 &  0.14  \\ 
$t_5 = t'_{BB''0}$  & $ \gamma_5/2$  &  $\gamma_5/2$ &  ---  &   ---      &    0.018          &  ---  &  --- &  ---   \\ 
$\Delta' = t'_{BB0}$    & $\Delta - \gamma_2 + \gamma_5$   &  $\Delta - \gamma_2 + \gamma_5$ & $\Delta - \gamma_2 + \gamma_5$  & 0.015  &      0.024     &  ---  & 0.018  &  0.022 \\  \hline 
  \end{tabular}
  \quad
\caption{
Comparison of the effective F1G0 effective tight-binding model (see text) 
parameters in the present work with corresponding values from the literature.   
Our results taken from Table \ref{table:effective} agree best with the graphite LDA data
of Ref. [\onlinecite{charlier}],
and are in reasonable agreement with experimental fits to Raman [\onlinecite{ramanbilayer}] 
and infrared data [\onlinecite{infraredfit,kuzmenko}] in bilayer graphene. 
Other tight-binding models for bilayer graphene 
motivated by graphite SMW parameters 
have been implemented with parameter sign differences 
in a variety of combinations leading to incorrect band structures.
\cite{partoens,latticepseudo,mccannrev,nilsson} 
Many properties of bilayer graphene are only weakly influenced by these remote hopping terms.
}
\label{table:comparison}
\end{table*}

Applications of the SWM model to generate tight-binding parameters for bilayer graphene
in the literature have been confused by  
variations in parameter signs \cite{partoens,latticepseudo,mccann,mccannrev,nilsson} 
which lead to inconsistent predictions for fine band features near the Dirac point.
(In Table \ref{table:comparison} we compare to values used in the 
literature for graphite and for bilayer graphene.)  
In an effort to clarify the confusion we define the effective velocity parameters  
 $\upsilon = -t_0 \sqrt{3}a/2\hbar$, $\upsilon_3 = t_3 \sqrt{3}a/2\hbar$ 
 and $\upsilon_4 = t_4 \sqrt{3}a/2\hbar$ so that all have positive values.  
 The LDA band effective velocities read from Table \ref{table:effective} are 
$\upsilon = 8.45\times 10^{5}$ m/s, $\upsilon_3 = 9.16 \times 10^{4}$ m/s and
$\upsilon_4 = 4.47 \times 10^{4}$ m/s.
 Recalling that the expansions near the Dirac points of the nearest neighbor structure factor 
 are (to linear order) 
 $f_1 ({\bf k}_D + {\bf q}) = \sqrt{3}a(-q_x + iq_y)/2\hbar$ and $f_1 (-{\bf k}_D + {\bf q}) = \sqrt{3}a(q_x + iq_y)/2\hbar$,
 where ${\bf k}_D = (4\pi/3a,0)$, 
 we rewrite Eq. (\ref{hamiltonian})
 using a notation similar to that in Ref. [\onlinecite{mccann,mccannrev}] by 
defining $\pi = \hbar( \xi q_x + iq_y) = \hbar q \exp(i \theta_{\bf q}) $ and 
$\pi^{\dag} = \hbar (\xi q_x - iq_y) = \hbar q \exp(-i \theta_{\bf q})$, 
where $\theta_{ \bf q}$ is a angle measured counterclockwise from the positive $x$-axis and 
 $\xi = \pm 1$  near $\pm {\bf k}_D$:
\begin{eqnarray}
{H} ({\bf k})     &=&  
\begin{pmatrix}
      0                      &      \upsilon  \pi^{\dagger}     &   - \upsilon_4 \pi^{\dagger}       &      - \upsilon_3 \pi     \\  
     \upsilon k \pi        &  \Delta'                 &      t_1    &  - \upsilon_4 \pi^{\dagger}     \\ 
    - \upsilon_4 \pi    &   t_1  &  \Delta'   &  \upsilon  \pi^{\dagger}  \\
    - \upsilon_3 \pi^{\dagger}      &      - \upsilon_4 \pi    &    \upsilon \pi       &     0    \\  
       \end{pmatrix}.
\label{conthamiltonian}
\end{eqnarray}
This Hamiltonian has a $-$ sign in front of the $\upsilon_3$ term when compared to Eq. (30)
of Ref. [\onlinecite{mccannrev}]. 
The correct choice of relative signs for weak hopping processes
can be relevant to the shape of the bands at low energies.
The negative sign of $t_0$ implies that intralayer bonding states have
lower energy than the antibonding counterparts.
The most relevant term for vertical tunnelling $t_1$ has a positive sign.
As shown in Fig. \ref{trigph} the relative sign of $t_1$ and $t_3$ determines 
the orientation of the trigonal distortion and the positions of the band-crossing points in bilayer graphene.
The correct sign choice ($t_1>0, t_3>0$) 
implies bands that are similar to those obtained with the incorrect choice 
$t_1<0, t_3<0$, \cite{latticepseudo,trilayergap}
whereas the incorrect mixed sign choice 
$t_1>0, t_3<0$ introduces a $60^{\circ}$ rotation.\cite{mccann,mccannrev}
Taking the relative signs of $t_0$ and $t_1$   
to be positive \cite{partoens,nilsson}
or negative\cite{mccann,mccannrev,latticepseudo,trilayergap}  
does not affect the trigonal warping orientation, but alters the way in which the 
$t_4$ term influences particle-hole symmetry breaking in the bands.  

These parameter sign issues also influence small terms in the  
$2\times2$  low energy Hamiltonian \cite{mccann,mccannrev} often used
to describe the low-energy bands.  Below, we rewrite
Eq. (38) of Ref. [\onlinecite{mccannrev}] (in the absence of an electric field between the layers)
with the sign of $t_3$ term corrected: 
\begin{eqnarray}
\label{twodimhamiltonian}
\hat{H}_{2} &=& 
 \hat{h}_{0} +  \hat{h}_{3} +  \hat{h}_{4} + \hat{h}_{\Delta} \\   
\hat{h}_{0} &=&-\frac{\upsilon^2}{t_1}
\begin{pmatrix}
0 & \left( \pi ^{\dag }\right) ^{2} \\
{\pi ^{2}} & 0
\end{pmatrix}, 
\nonumber \\
{\hat{h}}_{3} &=& -\upsilon_{3}
\begin{pmatrix}
0 & \pi  \\
\pi ^{\dag } & 0
\end{pmatrix}
+
\frac{\upsilon_3 a}{4\sqrt{3}\hbar} 
\begin{pmatrix}
0 & \left( \pi^{\dag }\right) ^{2} \\
-{\pi^2} & 0
\end{pmatrix}
 , \nonumber
\\
{\hat{h}}_{4} &=& \frac{2\upsilon \upsilon_4}{t_1}
\begin{pmatrix}
\pi^{\dagger}\pi & 0 \\
0 & \pi \pi^{\dagger}
\end{pmatrix}
,  \nonumber
\\
\hat{h}_{\Delta} &=& \frac{\Delta^{\prime} \upsilon^2}{t_1^{\prime 2}} 
\begin{pmatrix}
\pi^{\dagger}\pi & 0 \\
0 & \pi \pi^{\dagger}
\end{pmatrix}
.  \nonumber
\\
\nonumber
\end{eqnarray}
The effective mass parameter $m = t_1 / 2 \upsilon^2 $
takes the value of $\sim 0.044 m_e$ when calculated from LDA bands.
At $K$ ($\xi = 1$) retaining only the term linear in $p = \hbar q$ for the trigonal warping gives the eigenenergies
\begin{eqnarray}
E_{\pm} &=& \pm \left| \frac{ \upsilon^2}{t_1} p^2 \exp({-2 i \theta_{\bf q}}) + \upsilon_3 p \exp({i \theta_{\bf q}}) \right|   \nonumber \\
&=& \pm \left| \frac{ \upsilon^2}{t_1} p^2 + \upsilon_3 p \exp({i 3 \theta_{\bf q}}) \right|. 
\end{eqnarray}
From this expression we can see how the relative signs of $t_1$ and $t_3$ 
determine the orientation of the degeneracy-point triangle.
When $p =  \upsilon_3 t_1 / \upsilon^2$, the eigenvalues vanish for  
$\exp(i3\theta_{\bf q}) = -1$, {\em i.e.} when $\theta = \pi, \pm \pi/3$ as illustrated in 
Fig. \ref{trigph}.  
From the above equation we can also see that the sign of the particle-hole symmetry
breaking changes with the signs of $t_0$, $t_1$ and $t_4$ parameters.
Regarding particle-hole symmetry breaking we can see from 
Eqs. (\ref{twodimhamiltonian}) 
that the leading coefficient of the parabolic term $2 \hbar^2 \upsilon \upsilon_4 / t_{1} \sim 9 $ eV $\AA^2$ due to the $t_{4}$ 
is far greater than the intralayer second order expansion terms in Eq. (\ref{gfunc}) $C'_{AA,2}, C'_{BB,2} \sim -0.2$ eV $\AA^2$ from Table \ref{table:effective}
and indicates dominance of the interlayer coupling  in defining the particle-hole symmetry breaking
near the Dirac point.

\section{Summary and conclusions}
\label{sec:summary}
We have presented accurate tight-binding models for AB  
stacked bilayer graphene based on maximally localized Wannier functions that 
can capture the band structure in the entire Brillouin zone and near the Dirac point.
The models we have presented using orthogonal localized basis sets
are able to provide a different insight with respect to 
earlier tight-binding models for bilayer graphene  
that use non-orthogonal localized orbitals with a finite 
overlap between neighboring sites. 
Our full tight-binding model that includes
up to seventeen in-plane distant hopping terms 
is able to reproduce almost exactly the LDA bands in the whole Brillouin zone.
Effective models can be devised with fewer renormalised parameters
using fewer structure factors that provide simpler models that remain accurate 
near the Dirac points. This recipe can be used systematically to build tight-binding 
models with improved description of the bands in the whole Brillouin zone but using
fewer parameters.  We specifically describe a five parameter model analogous to the 
SWM model of graphene, and a model with fifteen paramters that includes up to $f_2$ and $g_2$ structure factors
and  offers an excellent compromise between simplicity and achieved accuracy.
Our {\em ab initio} approach has allowed us to assess the range of the 
microscopic hopping processes in graphene based systems.   
The parameters in SWM-type models are more correctly interpreted as effective hopping parameters 
that reproduce the bands near Dirac points. 

We note that LDA band parameters differ from phenomenological values in part because 
the experiments which determine the latter are often influenced by many-body effects.  
For example the effective intralayer nearest neighbour hopping parameter
is $\left| t_{AB1} \right| \sim 2.6$ eV for each graphene layer, 
similar to its value in single-layer graphene and corresponding 
to a velocity of $\upsilon \sim 0.84 \times 10^{6} $ m/s.
The value is $10 \% \sim 20 \%$ smaller than
$ \left| t_{AB1} \right| \sim 3 $ eV commonly extracted from
fits to experiments.  We ascribe this shift to 
the quasiparticle velocity renormalisation due to non-local exchange\cite{nonlocalexchange1,nonlocalexchange2,marginalfermiliquid}
that is not captured by the LDA.  
The strength of this renormalization depends on 
the carrier density and the dielectric environment,\cite{barlas,borghi,pomeranchuk} among other variable parameters, so it 
should be accounted for separately as necessary and not really be
incorporated into a universally applicable band structure model.  
At very low carrier density this physics can in principle lead 
to broken symmetry states both in bilayers \cite{latticepseudo,bilayerbs} and in 
ABC trilayers.\cite{trilayergap,trilayerbs}


The most important interlayer coupling effects are controlled by the  
effective $BA'$ coupling of $\sim 360$ meV
whose structure factor does not vanish near the Dirac point,
unlike the interlayer coupling mediated through terms 
linking $AA'/BB'$ with an effective strength of $\sim$ 140 meV each, 
that accounts for most of the particle-hole symmetry breaking near the Dirac points represented in Fig. \ref{trigph}, 
and terms coupling $AB'$ sites in the order of $\sim$ 280 meV responsible 
for trigonal warping.
The interlayer coupling leads to a small modification in the intralayer 
Hamiltonian that introduces a small site potential difference of $\sim$ 15 meV
between $A$ and $B$ (or equivalently $A'$ and $B'$ sites)
that influences the position of the higher energy $\pi$-bands.
The models we have presented provide a useful
reference for accurately modeling the first principles LDA 
electronic structure of bilayer graphene.

\begin{appendix}
\section{Tables for the parameters in the full tight-binding model.}
In this appendix we list in Tables \ref{table:full1}, \ref{table:full2} the distant tight-binding
hopping parameters used for AB stacked bilayer graphene,
associated with the $f_n({\bf k})$ and $g_{n}({\bf k})$ terms 
calculated using $30 \times 30$  $k$-point sampling density. 
\begin{table}[ht] 
\centering 
\begin{ruledtabular}
\begin{tabular}{|  c c c c  |  r r r | }  
\multicolumn{1}{|c}{ n}        &
\multicolumn{1}{c}{ m}        &
\multicolumn{1}{c}{ $N^0$}        &
\multicolumn{1}{c|}{ $d_n / a$}        &
\multicolumn{1}{c} { $t_{A B n}$ }    &
\multicolumn{1}{c} { $t_{A A^{\prime} n}$ }       &   
\multicolumn{1}{c|} { $t_{A B^{\prime} n}$ }     
  \\   \hline
  1   &  1  &   3   &  $\frac{1}{ \sqrt{3} }$        &   $-$3.010   & 0.09244       & 0.13912    \\
 2   &  3  &   3   &   $ \frac{2}{\sqrt{3}} $       &    $-$0.2387  & $-$0.01803  & $-$0.04753  \\
 3   &  4  &   6   &    $ \sqrt{\frac{7}{3}} $      &     0.01900  & $-$0.00068     &  $-$0.00108  \\ 
 4   &  7  &   6   &   $  \sqrt{ \frac{13}{3} } $  &    $-$0.01165  & 0.00181     &  0.00613  \\
 5   &  8  &   3   &   $ \frac{4}{\sqrt{3}} $       &    $-$0.01167 & 0.00029  & $-$0.00016  \\
 6   &  9  &   6   &    $  \sqrt{\frac{19}{3}} $   &    $-$0.00824 & $-$0.00019 & $-$0.00152  \\
 7   & 11 &    3  &   $\frac{5}{\sqrt{3}}$         &     0.00386  & $-$0.00079  &  $-$0.00163  \\
 8   & 13 &    6  &   $\sqrt{\frac{28}{3}}$       &    0.00250  & 0.00007   & $-$0.00152  \\
 9   & 14 &    6  &  $\sqrt{\frac{31}{3}}$        &    0.00224 & $-$0.00010  &  0.00075  \\
 10 & 16 &    6  &   $\sqrt{\frac{37}{3}}$       &    $-$0.00012  &  0.00052  & 0.00062  \\  
 \end{tabular}
 \end{ruledtabular}
 \caption{
Hopping amplitudes in eV units obtained in the  LDA calculations associated to $f_n({\bf k})$ 
structure factors connecting $AB$, $AB^{\prime}$ and $AA^{\prime}$.
}
 \label{table:full1}
\end{table}

\begin{table}
\begin{ruledtabular}
\begin{tabular}{|  c c c c |  r r r | } 
\multicolumn{1}{|c}{ n}        &
\multicolumn{1}{c}{ m}        &
\multicolumn{1}{c}{ $N^0$}        &
\multicolumn{1}{c|}{ $d_n / a$}        &
\multicolumn{1}{c} { $t_{A A n}^{\prime}$ }     &
\multicolumn{1}{c} { $t_{B B n}^{\prime}$ }       &
\multicolumn{1}{c|} { $t_{B A^{\prime} n}^{\prime}$ }       
  \\   \hline
   0 & 0 &   1   & 0                                           &    0.4295   &  0.4506  &  0.3310    \\
  1 & 2 &   6   & 1                                            &     0.22349 & 0.2260 &   $-$0.01016   \\ 
  2 & 5 &   3   &  $ \sqrt{3} $                            &    0.03692  &  0.03741 &  0.00049    \\
  2$^*$ &  $5^*$ &   3   &  $ \sqrt{3} $             &    0.03692 &  0.03741 & 0.00271   \\
  3 & 6 &   6   &    2                                          &   $-0.00253$  & $-$0.00163  &  0.00407  \\
  4 & 10 & 6   & $ \sqrt{7} $                              &   0.00076   & 0.00045 &  $-$0.00266  \\
  4$^*$ & $10^*$ & 6   & $ \sqrt{7} $                &   0.00076   &  0.00045 &  $-$0.00049  \\
  5 & 12 &    6  &  3                                           &   0.00327  & 0.00292  & $-$0.00180   \\
  6 & 15 &    3   &   $\frac{6}{\sqrt{3}}$              &   $ -$0.00085 & $-$0.00056  &  0.00222  \\
  6$^*$ & $15^*$ &    3   &   $\frac{6}{\sqrt{3}}$    &   $-$0.00085  & $-$0.00056  &  0.00015  \\
  7 & 17 &    6  &  $\sqrt{\frac{39}{3}} $                 &    $-$0.00031  & $-$0.00004  &   0.00143    \\ 
  7$^*$ & $17^*$ &    6  &  $\sqrt{\frac{39}{3}} $   &    $-$0.00031  & $-$0.00004   &   0.00034    \\
\end{tabular}
\end{ruledtabular}
\caption{
Hopping amplitudes in eV units obtained in the LDA calculations associated to $g_n({\bf k})$ structure factors 
connecting $AA$, $BB$ and $BA^{\prime}$.
We have introduced small equal shifts in the site terms $t'_{\alpha \alpha,0}$  
so that the $C'_{\alpha \alpha, 0}$ obtained from the truncated sum matches the 
values adopted for the position of the Fermi energy
in table \ref{table:effective}.
}
 \label{table:full2}
\end{table}

\section{Relationship between SWM and the effective tight-binding model parameters.}
In the following we paraphrase the analysis to compare the SWM model and the tight-binding model
for graphite presented in Ref. [\onlinecite{partoens}] particularised 
for bilayer graphene
and the simplified tight-binding model we proposed
in Eqs. (\ref{simpletbp1}-\ref{simpletb3}).
The SWM model for a bilayer graphene is simplified because
$\gamma_2 = \gamma_5 =0$ and the absence of the dispersion along the $z$-axis 
resulting in the Hamiltoinan
\begin{eqnarray}
{H_{SWM}}    
&=&  \begin{pmatrix}
 E_1                        &      0         &   H_{13}   &  H_{13}^*         \\  
    0                         &      E_2     &   H_{23}   &   -H_{23}^*    \\
    H_{13}^*             &    H_{23}^* &   E_3      &   H_{33}   \\
    H_{13}                &    -H_{23} &  H_{33}^*  & E_3      \\           
\end{pmatrix}
\label{hswm} 
\end{eqnarray}
where for a bilayer graphene $\Gamma=2$ we have
\begin{eqnarray}
E_1 &=& 2 \gamma_1 + \Delta \\
E_2 &=& -2 \gamma_1  + \Delta \\
E_3 &=&  0 \\
H_{13} &=& \frac{1}{\sqrt{2}} \left( - \gamma_0 + 2 \gamma_4 \right)  \sigma \exp (i \alpha) \\
H_{33} &=& 2 \gamma_3 \sigma \exp(i \alpha)
\end{eqnarray}
where $\sigma = \sqrt{3} a q / 2$ with $q$ measured from the Dirac point 
and $\alpha = \arctan(- q_x / q_y)$ is the angle measured from the $q_y$ axis.
We will compare with the eigenvalues obtained from the continuum Hamiltonian in Eq. (\ref{hamiltonian}).
Right at the Dirac point where $\sigma = 0$ and $F_{\alpha \beta}({\bf k}_D) = 0$ the four 
eigenvalues of the SWM Hamiltonian are $E_1$, $E_2$ and the doubly degenerate $E_3$.
Comparing with the four eigenvalues of the tight-binding Hamiltonian we can conclude that 
\begin{eqnarray}
\gamma_1 &=& t_1 \\
\Delta &=& \Delta'.  
\end{eqnarray}
The eigenvalues along the $\Gamma K$ line from the SWM bilayer graphene 
model are given by 
\begin{eqnarray}
E_{1,2} &=& \frac{\Delta}{2}  + \gamma_3 \sigma 
+ \gamma_1 \pm \frac{1}{2}\sqrt{C+D}  \label{firstswm}\\
E_{3,4} &=& \frac{\Delta}{2}  - \gamma_3 \sigma 
- \gamma_1 \pm \frac{1}{2}\sqrt{C+D}
\end{eqnarray}
where 
\begin{eqnarray}
C &=& \Delta^2 + 4 \gamma_3^2 \sigma^2 - 8 \gamma_3 \gamma_1 \sigma \\
&+& 4 \gamma_1^2 + 4 \gamma_0^2 \sigma^2 + 16 \gamma_4^2 \sigma^2 \\
D &=& 4 \gamma_1 \Delta - 4 \Delta \gamma_3 \sigma 
- 16 \gamma_0 \gamma_4 \sigma^2. \label{lastswm}
\end{eqnarray}

This is to be compared with the four eigenvalues of the effective 
continuum model whose eigenvalues are
\begin{eqnarray}
E'_{1,2} &=& \frac{\Delta'}{2} + t_{3} \sigma + t_{1} \pm
\frac{1}{2}\sqrt{C'+D'}  \label{firsttb}  \\
E'_{3,4} &=& \frac{\Delta'}{2} - t_{3} \sigma - t_{1} \pm
\frac{1}{2}\sqrt{C'-D'} 
\end{eqnarray}
where 
\begin{eqnarray}
C' &=& \Delta'^{2} + 4 \, t_{3}^2 \sigma^2 
- 8 \, t_{3} t_{1} \sigma  
\\
 &+&  4 \, t_{1}^{2} + 4 \, t_{0}^{2} \sigma^2 +16 \, t_{4}^{2} \sigma^2 \\
D' &=& 
 4 \, t_{1} \Delta
- 4 \Delta \, t_{3} \sigma
 + 16 \, t_{0}  t_{4} \sigma^2.  \label{lasttb}
\end{eqnarray}
When we compare the Eqs. (\ref{firstswm}-\ref{lastswm})
from SWM model and Eqs. (\ref{firsttb}-\ref{lasttb}) formally there is only one difference
in the sign of the last term of Eq. (\ref{lasttb}) that would lead to 
$-\gamma_0 \gamma_4 = t_0 t_4$ if we want 
all the eigenvalues to remain the same.
The physical choice of signs consistent with our calculations is 
$t_0 = -\gamma_0$ and $t_4 = \gamma_4$.

\end{appendix}

\begin{acknowledgments}
This work was supported by the US Department of Energy
Division of Materials Sciences and Engineering under grant DE-FG03-02ER45958,
by Welch Foundation grant TBF1473, and by the National Science Foundation under Grant No. DMR-1122603.
JJ was partially supported by the National Research Foundation of Singapore under its Fellowship program (NRF-NRFF2012-01).
We gratefully acknowledge the assistance and the computational resources provided by 
the Texas Advanced Computing Center.
We thank the developers of wannier90 and Quantum Espresso for making 
these tools freely available.
\end{acknowledgments}


\begin{thebibliography}{99}
\bibitem{footnote}
In single-layer graphene the coupling between $\sigma$ and $\pi$ bands
vanishes by mirror-plane symmetry.  Coupling is present in bilayers, although the 
hopping amplitude between a $p_z$-orbitals at a carbon site 
and a neighboring bonding $\sigma$ orbital is smaller than 0.1 eV in magnitude. 
Considering the large separation in energy over 10 eV between the $p_z$ and the $\sigma$ 
orbitals this weak coupling implies negligibly small mixing between the two types of orbitals.
\end{thebibliography}
\end{document}